# Fabrication of High-Aspect Ratio Nanogratings for Phase-based X-ray Imaging

Martyna Michalska, Alessandro Rossi, Gašper Kokot, Callum M. Macdonald, Silvia Cipiccia, Peter R.T. Munro, Alessandro Olivo, and Ioannis Papakonstantinou*

**M. Michalska, A. Rossi, I. Papakonstantinou**: Photonic Innovations Lab, Department of Electronic & Electrical Engineering, University College London, Torrington Place, London WC1E 7JE, UK.

**G. Kokot**: Jožef Stefan Institute, Jamova cesta 39, SI-1000 Ljubljana, Slovenia

**C.M. Macdonald, S. Cipiccia, P.R.T. Munro, A. Olivo**: Department of Medical Physics and Biomedical Engineering, University College London, Malet Place, Gower St, WC1E 6BT, UK

*i.papakonstantinou@ucl.ac.uk

**Abstract:** Diffractive optical elements such as periodic gratings are fundamental devices in X-ray imaging – a technique that medical, material science and security scans rely upon. Fabrication of such structures with high aspect ratios at the nanoscale creates opportunities to further advance such applications, especially in terms of relaxing X-ray source coherence requirements. This is because typical grating-based X-ray phase imaging techniques (e.g., Talbot self-imaging) require a coherence length of at least one grating period and ideally longer. In this paper, the fabrication challenges in achieving high aspect-ratio nanogratings filled with gold are addressed by a combination of laser interference and nanoimprint lithography, physical vapor deposition, metal assisted chemical etching (MACE), and electroplating. This relatively simple and cost-efficient approach is unlocked by an innovative post-MACE drying step with hexamethyldisilazane, which effectively minimizes the stiction of the nanostructures. The theoretical limits of the approach are discussed and, experimentally, X-ray nanogratings with aspect ratios >40 demonstrated. Finally, their excellent diffractive abilities are shown when exposed to a hard (12.2 keV) monochromatic x-ray beam at a synchrotron facility, and thus potential applicability in phase-based X-ray imaging.

## 1. Introduction

The ability to fabricate high aspect ratio silicon nanostructures is fundamental in many applications spanning sensors,[1] batteries,[2] solar cells,[3] superhydrophobic[4,5] and mechano-bactericidal materials,[6] as well as X-ray imaging.[7,8] In the latter, diffractive optics such as periodic gratings are one of the key tools used in phase-based approaches. Here, nanostructuring opens new avenues to further advance phase-based X-ray imaging utilizing gratings, since the exploitation of coherent phenomena such as Talbot self-imaging requires the X-ray illumination to have a coherence length at least equal to, and ideally larger than, a grating period. Furthermore, the recent discovery of the infinite moiré effect opens new avenues to perform phase-based imaging with increased sensitivity and simplified setups.[9,10]

A key parameter to consider in order to perform this type of imaging is the need to attain a sufficiently large phase shift (typically $\pi$ or $\pi/2$) at the desired X-ray energy, with higher energies ("harder" X-rays) providing access to a wider range of applications, but at the same time requiring taller structures to obtain a certain phase shift. Combined with periods of the order of hundreds of nanometers, this leads to high aspect ratios (>10, often much larger), that can be reduced only by employing higher Z materials such as gold.[11] Concurrently, characteristics such as vertical profile for the trenches, regularity of the duty cycle and uniformity across the sample need to be taken into account. For this first proof-of-concept test we are targeting energies of the order of 12 keV widely used in e.g., synchrotron research; in the future, we will strive to develop thicker structures in order to access a wider range of application, e.g., medical imaging, which starts at ~17.5 keV for mammography and requires energies of up to 100 keV in other, more challenging areas.

To accommodate these constraints various techniques have been employed that predominantly rely on creating a polymer or silicon template with a conductive seed layer at the bottom of the trenches or conformally deposited along them for the subsequent electroplating of Au. The location of the seed layer dictates the metal deposition mechanism, which follows a bottom-up growth or from the sidewalls, respectively.

Although electroplating on polymer templates has the benefit of inducing a desirable void-free bottom-up metal growth due to an insulating character of polymer walls, the attainable aspect ratio is limited. Hence, the focus has turned to fabricating silicon templates via either deep reactive ion etching (DRIE) or metal assisted chemical etching (MACE). Since vertical side walls are a prerequisite, cryogenic DRIE is typically needed to avoid scalloping and etch taper under standard dry etching conditions, which in turn requires specialized equipment unlike MACE. The latter, first reported by Li and Bohn,[12] is less limited in terms of aspect ratio due to the highly localized electrochemical reaction occurring on silicon-metal catalyst interface in addition to being facile, cost-efficient, and scalable.[7] Broadly, it relies on patterning of a noble metal (Au, Ag, Pt) on silicon which acts as both etch mask and catalyst to reduce hydrogen peroxide ($H_2O_2$) in a an etchant solution (hydrofluoric acid (HF) and water). During the oxygen reduction, electron holes are generated and injected into the underlaying silicon oxidizing it to $SiO_2$, which is subsequently etched by HF making the metal catalyst sink down. To date, the technique has been largely explored, particularly, in terms of etch rates, porosity, and etch profile, which can be tuned by substrate choice (resistivity, doping type), catalyst (type, porosity), $H_2O_2$/HF ratio, alcohol additives, temperature, supporting structures etc.[8,11,13] However, there are still two main challenges when the reactions need to be carried out at the nanoscale pitch and high aspect ratio. These include 1) stiction and agglomeration of structures during drying step, and 2) taking advantage of a bottom-up metal electroplating from the existing seed layer.

The first issue occurs with an increasing grating height, and it is dependent on the mechanical properties of the used material and the forces acting upon it.[14] There is a maximal lamella height before an irreversible collapse occurs, mainly attributed in the literature to van der Waals forces, as well as capillary forces between the silicon surface and liquid-air interface during the drying process.[15] A prevention of collapse behavior is fundamental in preserving pattern-induced properties (optical etc.) as well as allowing for a conformal coating or gold filling. Until now, the issue has been solved by either employing sophisticated post etching drying steps (critical point drying (CPD),[16] freeze drying[17]) or adding transversal supporting structures such as bridges.[18] The former requires special equipment and has limited scalability (CPD in particular). The latter complicates the design so that more complex methods are required to create a pattern in the first place, such as E-beam lithography as opposed to simple laser interference lithography (LIL). Alternatively, MACE in the gas phase has been demonstrated to work,[7] albeit again requiring a bespoke setup instead of an ordinary wet bench, which is one of the aspects making this etching approach so attractive.

It appears appealing to utilize the remaining post-MACE catalyst uniquely located at the bottom of the trenches as a conductive seed layer for the subsequent bottom-up electroplating. However, in practice, this is not easy to achieve, and a mixed growth is often favored. This likely originates from the semiconducting character of silicon walls and the presence of some level of porosity resulting from the etching process itself, which acts as nucleation sites.[19] To attain insulating walls, at least 150 nm of oxide layer is required to ensure large enough electrical resistivity[20], which puts a limit on the pitch of the grating. For a larger pitch, on the other hand, this can be achieved via e.g., thermal oxidation. Yet, a change in a duty cycle needs to be factored in the design due to the nature of silicon oxidation itself. Some additives in the electrolyte may be incorporated to promote preferential plating at the bottom.[21,22] Alternatively, one can use chemical or physical vapor deposition (CVD or PVD, respectively) to conformally coat the walls with a seed layer and therefore induce the growth from the sidewalls towards the middle of the trenches. Within CVD approaches, atomic layer deposition (ALD) is unrivaled to accommodate any pattern resolution, and it has been shown by Miao at al.[23] to work for the nanogratings of 200 nm period, which were ultimately filled with Pt. Although PVD is thought to be less suitable for nanogratings, we show here that gratings with a half-pitch of ~200 nm can be successfully sputtered to achieve electrical continuity. We find this to be strongly dependent on the quality of the adhesive layer. It is noteworthy that PVD is significantly cheaper and easier to implement within the process, and the entire step takes 1.5 h instead of several hours for a typical ALD step. Additionally, highly uniform sidewalls-growth can serve for halving the pitch in a similar manner as iterated spacer lithography.[24,25]

In this paper, we present a simple fabrication method, which consists of laser interference and nanoimprint lithography, MACE, and electroplating, for X-ray nanogratings with high aspect ratios (>40). This is enabled by our post-MACE drying step, which effectively minimizes the stiction of the nanostructures. In the electron microscopy community, evaporation of a low surface tension hexamethyldisilazane (HMDS) is a known drying alternative to CPD, used to preserve morphology of biological specimens.[26] Here, we successfully adapt it to silicon and show experimentally ~2-fold improvement in preventing the lamellae from collapsing when compared to drying from water, thus achieving ~7 µm tall, defect-free nanogratings. We interpret it is the HMDS chemical ability to lower surface energy of the gratings that also contributes to the effect. However, as we will discuss later, we observe a pronounced discrepancy between the theory predictions for maximal grating height and experimental observations. We identify this divergence is in part caused by the incorrect assumption of the grating deflection axis and derive a condition for when the corner of the lamella will deflect first. Nevertheless, this approach allows to maintain the simplicity of the design and processing steps by being fast (minutes) and removing the need for special tools. While it can impact other fields where the collapse behavior constitutes a problem, here it allows us for a successful deposition of a conformal gold seed layer via sputtering and high-density gold filling of the trenches. Alongside detailed characterization via electron microscopy and other methods demonstrating that the design goals have been achieved, we also exposed one of the fabricated gratings to a hard (12.2 keV) monochromatic x-ray beam at a synchrotron facility and used the resulting diffracted powers to gain additional insight on the gratings' structure and their future usability as imaging devices.

## 2. Results and discussion

Figure 1 demonstrates fabrication steps, which, broadly speaking, include a generation of linear grating pattern (1-2), its transfer into silicon by MACE (3), and gold electroplating from a sputtered seed layer (4-5).

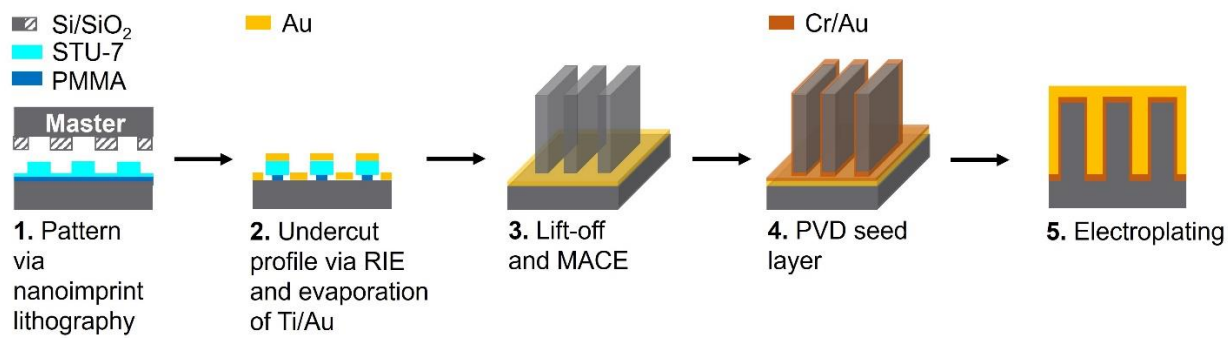

**Figure 1.** Overview of the nanogratings' fabrication. (1) The process begins with pattern generation via nanoimprint lithography (NIL) by stamping a master (prepared by laser interference lithography (LIL)) onto a polymer bilayer. (2) The bi-layer facilitates an undercut profile, which is formed by reactive ion etching (RIE), followed by e-beam evaporation of a metal catalyst. (3) After the lift-off process, which defines an etch mask, metal assisted chemical etching (MACE) is performed yielding high aspect ratio nanogratings. (4-5) Finally, a conformal conductive seed layer is sputtered onto the grating allowing to subsequently fill the trenches with gold via electroplating.

Specifically, the process begins with a master preparation, followed by its replication via nanoimprint lithography (NIL; Figure 1, step 1). To prepare the master, we pattern a photoresist on a silicon wafer with a $SiO_2$ layer of 300 nm by means of LIL using a one-mirror Lloyd's interferometer setup (Figure S1). Subsequently, we transfer the pattern into glass by reactive ion etching (RIE) and $CHF_3$/Ar plasma with varying $H_2$ content. The latter allows for a convenient tuning of the grating width and, in turn, of the duty cycle as we recently reported[27] – the higher the $H_2$ flux, the larger the duty cycle (Figure S1).

We then use NIL to replicate the master and prepare a series of patterns for the subsequent etching (Figure 1, step 2). Such pattern generation is cost-effective, fast and scalable; when done on a bilayer, it also allows lift-off.[28] Here, we stamp the master on the bilayer consisting of PMMA sacrificial layer and STU-7 NIL resist (Figure 2a). The thickness of both layers needs to be optimized so that both residual layers are minimal to prevent a prolonged oxygen breakthrough, which would otherwise lead to a significant change in duty cycle (Figure S2). Note that the thickness of STU-7 can be precisely calculated by knowing the pattern volume of the master. We achieve undercut by dry etching in oxygen plasma and taking advantage of the difference in etching rates between the layers and optimizing the etching time (Figure 2b-c). The inset in Figure 2c presents the resulting gold pattern (2 nm Ti and 15 Au) after electron beam evaporation and lift-off.

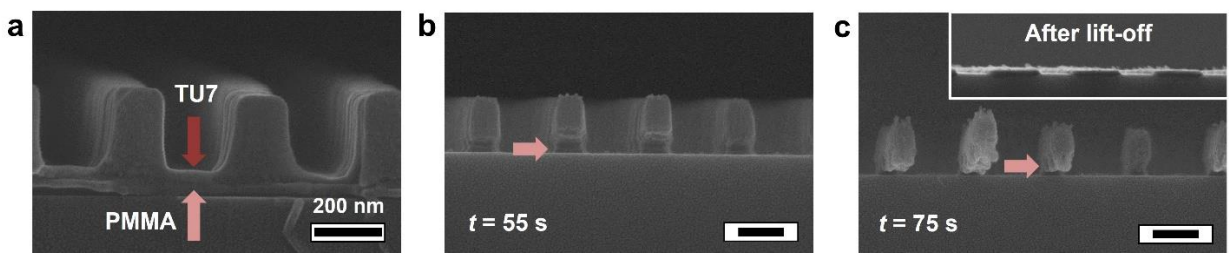

**Figure 2.** Bi-layer lift-off process via nanoimprint lithography. SEM images of PMMA (bottom layer) and STU-7 (TU7; upper layer) nanoimprinted layer before (a) and after the sacrafical layer descumming for 55 and 75 s (b and c, respectively). An inset in (c) shows 15 nm Au catalyst on 2 nm Ti adehsive layer on silicon, which were electron beam-evaporated and lifted-off, in order to define the etch mask for the subsequent MACE step.

In the next step (Figure 1, step 3), we perform MACE at standard conditions as previously described (see Methods),[8] while carefully controlling the temperature, which we find to be the most sensitive parameter to permit vertical etch, and thus features of high aspect ratio. Hence, the etching is conducted in an ice bath (~7 °C). This slows down the transport and, crucially, it prevents catalyst mobility at the expense, however, of the etching rate (ca. 45 nm/min). Note that catalyst movements such as rotation and folding can lead to interesting patterns,[29] albeit limiting the attainable aspect ratio because of the additional strain introduced by such movements. Figure 3a presents a defect-free nanograting of height $L = 3.6$ µm. With an increase in the etching time, however, we reach a height limit, above which the stiction between lamellae occurs during the drying step, deteriorating the pattern (Figure 3b).

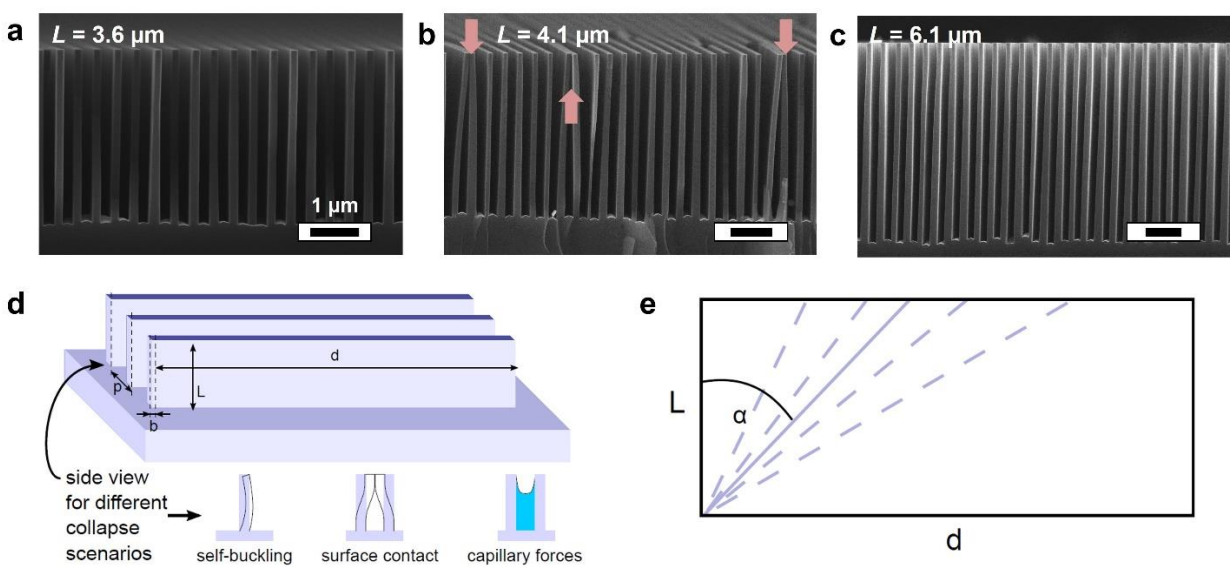

**Figure 3.** The solvent effect after the post-MACE drying step on the gratings' integrity. SEM images of nanogratings after MACE, dried directly from water (a-b) and using our HMDS approach (c). Scale bars 1 µm. (d) A schematic of the grating and its dimensions used for modelling. The deflection is assumed along $d$ axis. (e) A shematic demonstrating the corner bend

with several possible deflection axes (purple, dashed). The axis leading to minimal elastic restoring force is drawn as a full purple line and the α angle from Equation 1 is indicated.

Having assumed that capillary action is a driving force of the collapse behavior, we hypothesize that drying by evaporation of a low surface tension fluid like HMDS could minimize the effect. After all, HMDS is successfully used in the preparation of biological samples for electron microscopy. There, it relies on fixing the specimen in glutaraldehyde, its gradual dehydration in increasing ethanol (EtOH) concentrations, and immersion in HMDS, followed by air-drying. Remarkably, transferring the nanograting directly from water into 50% and 100% EtOH, followed by dipping in HMDS, and allowing it to air-dry, preserves the vertical profile and pattern integrity as shown in Figure 3c. Etching for 2.5 h leads to gratings of ~7 μm height with well-preserved patterns that we will use for further investigations.

To elucidate the underlying mechanism, we vary the drying conditions (Table 1, samples 1-4) to find out that both ethanol and HMDS are required as an intermediate and final solvent, respectively, to achieve the desired effect. We assess the effect based on SEM imaging of the cross sections and top views across the entire surface, as well as by a visual inspection. As expected, unlike biological specimens necessitating penetration through a mesh of their internal and fragile structures, here one ethanol concentration is sufficient to replace water, due to the solid nature of silicon. A direct transfer of the sample from water to HMDS (sample 4) leads to an observable phase separation on the drying surface, due to the high immiscibility of the liquids. Whilst the HMDS phase dries instantly, the water phase requires several hours and leaves a distorted pattern behind, assessed by SEM top view micrographs (Figure S3). The pattern distortion changes the optical properties of the sample, and the collapsed areas can be noted upon a visual inspection and clearly distinguished from the defect-free diffractive regions (Figure S3, light and dark regions).

Interestingly, despite EtOH and HMDS having similar surface tensions (22.4 and 18.2 mN/m, respectively), drying only from EtOH (sample 3) leads to the structural collapse (Figure S4). These experimental results suggest capillary forces are indeed an important piece of the puzzle, although as we demonstrate not in a sense typically considered in existing calculation approaches.

**Table 1.** Summary of the drying protocols.

| Sample ID | H$_2$O | EtOH 50% | EtOH 100% | HMDS | Effect |
|---|---|---|---|---|---|
| 1 | x | x | x | x | defect-free |
| 2 | x | - | x | x | defect-free |
| 3 | x | - | x | - | collapse |
| 4 | x | - | - | x | collapse |

HMDS is routinely employed in cleanroom processes as an adhesion promoter prior to the photoresist coating. On a water-free surface, it works by its ability to chemically bond its silicon atom to the oxygen of oxidized surface. After the HF etching, which removes SiO$_2$, we expect the predominant coverage of the surface with hydride group with however, a small percentage of hydroxylation due to the presence of water and hydrogen peroxide.[30] We compare water contact angles $\theta$ on the non-patterned silicon surface before and after MACE (sample nitrogen-dried directly from water) and we note a significant increase in $\theta$ (Figure S5), which confirms the domination of hydride groups. After drying the sample with HMDS, $\theta$ increases further (from 48 to 87°), indicating chemically bonded molecules, and hence the presence of some hydroxylation induced during the etching. We also measure the water contact angle on grating 2 (Table 1) and obtain $\theta = 147°$.

The hydrophobic character of the gratings after the HMDS treatment is sufficiently robust to prevent inducing the collapse behavior after sonicating samples in water (15 min; the sample floats on the water-air interface). Repeating the same test with EtOH results in collapsed edges of the grating, albeit a careful inspection across the sample reveals a well-preserved pattern (Figure S6). Upon removing the silane in oxygen plasma, which also does not damage the pattern, collapse can be induced by placing the sample in water. This further validates that a change in surface energy plays a role in protecting the pattern.

We now compare our experimental results with theoretical models that predict collapse behavior to better understand the underlying mechanism as well as the limits of our approach, and how to potentially overcome them. Such limit can be expressed as the maximal lamella height, $L_{max}$, before an irreversible collapse occurs. The particular force responsible for the collapse remains elusive as we confirm the standard calculations greatly overvalue $L_{max}$, though we make important steps towards understanding how exactly the lamella collapses based only on restoring elastic force arguments.

We consider a cuboid lamella of width $b$ = 80-250 nm (matching our experiment), and length, $d$ = 1 cm (Figure 3d). We begin our analysis (see SI for details) with the traditional assumption that deflection of a cuboid happens along $d$ and in the direction of the smallest dimension, which in our case is $b$. We then follow several scenarios for forces that could lead to buckling.

We start with self-buckling (*SB*) – a collapse initiated solely by gravity acting on the lamella – and find that the prediction from the Greenhill result of $L_{maxSB}$ = 5.3 mm vastly overestimates the experimental value $L_{max}$ < 10 µm, even before we introduced the HMDS method. This large discrepancy has been consistently found in experiments with distinct protrusion geometries.[31,32] Figure S7 shows the achieved, albeit not ultimate, experimental limits of the aspect ratio of nanogratings after introducing HMDS-mediated drying. While there is no doubt that 7.2 µm-tall gratings (aspect ratio 43) can be achieved, the images of 12.2 µm-tall patterns (aspect ratio 117) imply that the collapse here is more likely driven by our inability to control the etching anisotropy (Figure S7b). Multiple regions exhibit defect-free characteristics as demonstrated in Figure S7c-d. Although this obstacle prevents us from showing a true limit for our drying approach (hence $L_{max}$ with inequality sign), it does not prevent us from further discussing the theory of the underlying mechanism.

With self-buckling ruled out, we now consider scenarios where a force bends two lamellae until they come into contact. A good candidate are capillary forces, because we experimentally know the collapse happens during the drying process. We compare the capillary force for our geometry, where we make the usual simplification of two infinite parallel lamellae, to the elastic restoring force and calculate $L_{max}$ for capillary force, $L_{maxCF}$ = 107 µm. This result reveals that the capillary force, as obtained with the underlying assumptions for two infinite parallel lamellae, is not large enough to bring them in contact at the measured $L_{max}$.

Surface interaction is another candidate to explain the lamella collapse. Following the result from Hui *at al.*[32] and using our notation for geometrical parameters, we obtain an $L_{max}$ for surface energy interaction, $L_{maxSE}$ = 2 µm, which is indeed less than the experimentally determined $L_{max}$. The computed $L_{maxSE}$ can be larger, if one is to include electrostatic interaction, as has been shown for pillars.[14] As discussed earlier the effect of surface energy modification by HMDS can influence the exact number $L_{maxSE}$. Yet, we cannot brush off the fact that the surface interaction scenario comes with the surfaces already in contact. However, the result does assure us that any lamella height above $L_{maxSE}$ will result in a collapse, if we apply a force great enough to bend them towards each other.

Another suggestion for a lateral force on the lamella is the Laplace pressure variation due to surface roughness and lamella thickness variation. As has been pointed out before, the Laplace pressure differences do not exist if the liquid between each lamella space is not strictly separated (as would be the case for infinitely long lamellae). For lamellae with finite length, during the drying, the liquid stays connected at the bottom edges outside of the grating. Nevertheless, variations in $b$ and defects present in experiments can influence the local bending condition.

Despite having already established that capillary forces between two infinite parallel lamellae are not responsible for the collapse, we have experimental evidence suggesting capillary action is not to be discarded. Figure S6 shows a grating with $L$ close to $L_{max}$, where the lamella deformed at the corners and is still standing unbent in the central part. The deformation does not happen around the axis assumed for the infinite parallel lamellae (Figure 3d) but rather a differently oriented, much shorter axis. We calculate the elastic restoring force for the lamella corner, compare it to the same type of force for the lamella cuboid and derive a condition for lamella length ($d$) at which the bending of the corner is preferable (i.e., the force needed to bend it is the smallest, (Figure 3e)):

$$d > \frac{4\,L^3}{3\sin(2\alpha)\delta^2}, \text{(Eq. 1)}$$

Where $\alpha$ is the angle between the vertical lamella edge and the axis of corner buckling (see Figure 3e) and $\delta$ is the deflection, which in our case is half the pitch ($\delta = h/2$). The right side of Eq. 1 is minimized at $\alpha = \pi/4$, which predicts the optimal buckling angle for the corner. In the range of our experimental values for $L$, $\delta$ and $d$, Eq.1 is always fulfilled, thus predicting that corner collapse happens first. The corner of the lamella is a strong pinning point for the liquid interface, and we experimentally influence $L_{max}$ by changing the drying agent. Unfortunately, an analytical expression for a capillary force in a situation described here is lacking in the literature and is beyond the scope of this paper. However, a scenario, where the lamellae touch their surface first at the corners of the grating and then continue the collapse by mechanism of surface energy is supported by our analysis.

After successful fabrication of the vertical grating, we now fill the space between the trenches with gold in order to prepare it for the subsequent evaluation of the X-ray performance. Silicon is the material of choice for gratings fabrication due to its very good robustness, its relative X-ray transparency, and low cost. A conductive layer (e.g., Au, Pt, Cu) reduces the overpotential necessary to allow plating, driving the nucleation and growth of metal onto itself. The quality and nature of the seed layer, combined with the plating method, can change the electrochemical deposition in terms of plating rate, morphology, and density.[33,34] To ensure good electrical contact and prevent any mechanical damage during handling the nanoscale gratings, we introduce a frame surrounding the design (Figure S9). This proves to be of paramount importance at such scales for certain geometries (see Supplementary Text 2, Figure S10).

We first attempt an electroplating on the remaining post-MACE gold catalyst layer, which is conveniently located at the bottom of trenches. Unfortunately, this approach led to a mixed growth, likely due to some porosity of the sidewalls. We therefore follow a different approach to force an electrochemical deposition in a conformal way by depositing a seed layer of the same material (Au) via sputtering. We demonstrate that high-pressure sputtering can successfully deposit a quality seed layer on high-aspect ratio nanogratings (Figure 4a-c). Here, we find that an optimal thickness of adhesive layer (Cr in this instance) is key to achieve an electrical continuity across the grating. An excessively thin the layer leads to a patchiness in the electroplated gold, which follows the areas where a continuous seed layer was present (Figure S8a-c). Doubling the time of Cr deposition while maintaining the Au sputtering conditions (7.5 min) results in a full coverage along the sidewalls with a mosaic structure (Figure 4c, Figure S8d), similar to the one obtained by ALD.[9] Finally, we successfully electroplate the gratings across a full length of the trench (Figure 4d-f, Figure S8e-f) by using a sulphite-based gold plating solution and a direct cathodic current (see Methods).

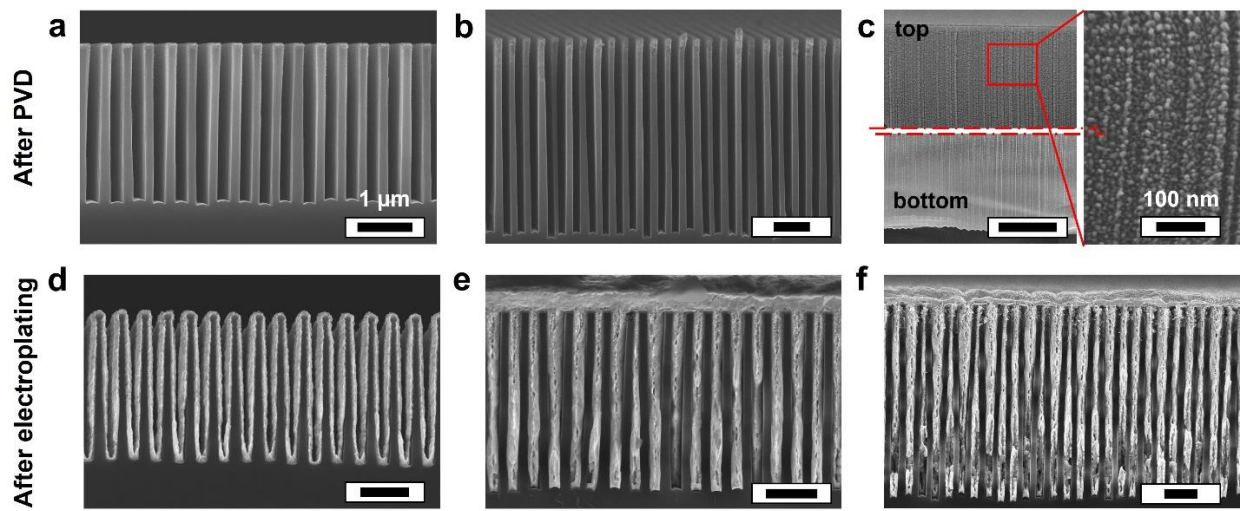

**Figure 4.** A seed layer on silicon mold and gold-filled trenches by electroplating. (a-c) SEM cross sectional images showing the Cr/Au seed layer on high-aspect ratio nanogratings and (d-f) the results of gold electroplating. The seed layer was succesfully sputtered on gratings with two heights: (a) $L = 3.2$ μm and (b) $L = 6.1$ μm. (c) A magnification of a face-view of the lamella confirms the seed layer deposition and reveals its granular structure ($L = 6.1$ μm). (d) A conformally electroplated gold following the silicon mold can be utilized for halving the period. (e-f) Trenches of high-aspect ratio silicon mold fully filled with gold. Aspect ratios are 16 (e) and 30 (f). Scale bars are 1 μm, unless stated otherwise.

The design of the gratings is optimized to diffract hard X-rays with low absorbance, therefore one of them was tested at the I13-1 beamline of the Diamond Light Source (United Kingdom) for its diffractive ability. We tested a grating with a 414 nm period, a ~50% duty cycle, height of $3.2 \pm 0.2$ μm, and an estimated Au density of 87%. The sample was placed at 3.3 m from a detector (see

Methods) and exposed to a monochromatic beam with energy of 12.2 keV, with a pre-collimating beam aperture width of 440 µm in the direction orthogonal to the grating trenches. The diffraction orders' intensities are a quantitative measurement of the grating performance, and they depend on the phase modulation induced by the Si and Au lamellae. The total thickness of the lamella and the Si substrate is ca. 200 µm, which can cause a non-negligible attenuation of the beam at the used energy. Although this can be reduced by using a higher energy, the transmitted intensity was sufficient to quantitatively estimate the diffracted powers.

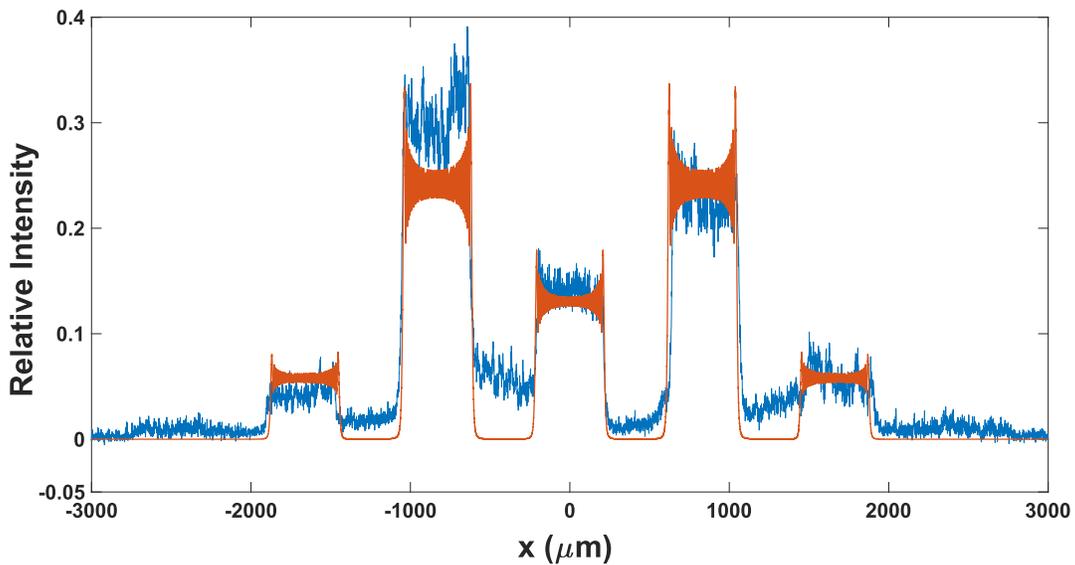

**Figure 5.** The intensity profile of the diffracted X-ray beam. The blue line and the orange lines show the experimental data and the simulated signal, respectively, for a 12.2 keV beam diffracted by a 414 nm-period and 3.2 µm tall grating with a gold density of 87%.

The simulated intensity profile (Figure 5, orange line) largely agrees with the experiment (blue line), providing further confirmation that the design specifications were met. The grating successfully splits the beam into several visible diffraction orders, which we collected up to the 2$^{nd}$ order. The 1$^{st}$ orders collect the highest intensity, equal to approximately 24% of the transmitted flux. The asymmetry between the 1st diffraction orders may be attributed to some tapering and catalyst mobility slightly varying lamellae shape.[23] Both can be corrected by fine-tuning oxidant/etchant ratio and more efficient temperature control by mixing (see Methods), respectively.[35,36] Likewise the excess intensity between diffracted orders can be attributed to residual grating imperfections and especially to x-ray scattering, notably by air considering there was a 3.3 grating-to-detector distance in the used setup, without an evacuated tube.

## 3. Conclusion

Attaining high-aspect ratio, dense nanostructures is a long-standing challenge being systematically addressed by advances in fabrication processes. Our method shows significant improvements in respect to the state-of-the-art in fabrication of nanogratings, which typically requires expensive and time-consuming steps such as DRIE for etching or ALD to introduce seed layer. In this paper, instead, the sputtering gives a satisfactory seed layer coating and can be performed within 120 min. Furthermore, to the best of our knowledge, we demonstrate for the first time a successful etching of high-aspect ratio sub-micron gratings without additional supporting structures via wet MACE. We achieve it by our new drying step based on HMDS that replaces more sophisticated processes. Finally, the fabricated gratings possess the expected diffractive abilities when exposed to a monochromatic x-ray beam, with most of the transmitted flux being redirected to higher order energies. This is a fundamental property for interferometric phase contrast imaging and the sub-micron period of the grating can extend its applicability to low coherence compact sources and, ultimately, find applications in applications across the life and physical sciences.

## 4. Experimental Section

The fabrication process consists of (1) master preparation via laser interference lithography (LIL) and glass etching, (2) master replication via nanoimprint lithography (NIL), (3) metal assisted chemical etching (MACE), and (4) electrodeposition of gold (Figure 1).

*Master preparation:* Master with a linear grating of period $p$ = 400 nm was prepared via LIL using a Lloyd's mirror interferometer setup as we previously reported.[25] First, a silicon wafer (p-type, boron-doped, <100>, 1-10 Ohm cm, 675 μm thick; MicroChemicals) with 300 nm of $SiO_2$ layer was cleaned with acetone and isopropyl alcohol (IPA). A photoresist ma-N 400 (Microresist Technology) was diluted 2:1 with 4-methyl 2-pentanone, spin coated onto the wafer at 4000 rpm, and baked at 100 °C for 2 min. Subsequently, the wafer was illuminated at 24° for 90 s using the free-space UV laser with a beam diameter of ~1 mm (IK3201R-F by Kimmon; Class 3B; 25 mW; 325 nm; CW; single mode TEM00). The pattern was then developed using AZ 726 MIF (5:1 dilution with deionized water; Microchemicals) for 2 min. An oxygen descum etch was performed to remove any residual photoresist. The pattern was registered in glass by reactive ion etching (PlasmaPro NGP80 RIE, Oxford instruments) under the following conditions: $CHF_3$ 15 sccm (standard cubic centimeters per minute), Ar 50 sccm, 215 W, 30 mTorr, and 20 °C. To increase the grating width, $H_2$ was used at 6 sccm.

*Nanoimprint lithography:* To replicate the pattern for a subsequent lift-off process, which requires an undercut profile, NIL was performed on a bi-layer deposited on silicon. First, a silicon wafer (p-type, boron-doped, <100>, 1-10 Ohm cm, double polished, 200 μm thick; MicroChemicals) was cleaned as above, and subsequently spin coated with 1) poly(methyl methacrylate) (PMMA 950 A2; MicroChem Corp.) at 2000 rpm and baked at 160 °C for 5 min, followed by 2) STU-7 (Obducat) and baked at 100 °C for 1 min. Depending on the master profile,

STU-7 was diluted accordingly to achieve a minimal residual layer after the imprint. The NIL process was performed on EITRE 3 (Obducat). The master was first replicated into an intermediate polymer stamp (IPS; Obducat) at 155 °C and 30 bar for 20 s. Transfer of the negative pattern into the STU-7 was operated at 70 °C and 40 bar for 60 s with 120 s of UV curing. The undercut profile was achieved by $O_2$ RIE (20 sccm, 50 W, 50 mTorr) where PMMA layer was intentionally over-etched (75 s).

*Metal assisted chemical etching:* First, the metal catalyst (2 nm Ti and 15 Au) was electron beam evaporated onto the pattern with deposition rate of 0.1 nm min$^{-1}$. Lift-off was performed in dimethyl sulfoxide (DMSO) via sonication at 65 °C for 5 min twice. Prior to the MACE, samples were subjected to $O_2$ plasma for 1 min, which increased the etching rate 3-fold. The MACE was carried out in a solution comprised of 5.3 M HF, 0.25 M $H_2O_2$ and 50 M $H_2O$ at ca. 7 °C for 1-5 h, and then rinsed thoroughly with $H_2O$ and stored in $H_2O$ until the drying step. To achieve and maintain the low temperature, a beaker containing the samples was placed in an ice bath. The ice was replenished every 20 min and a manual mixing was employed every 5 min. Due to technical constraints when handling HF, other mixing methods were unable to be implemented, which we plan to address for the future processes. The samples were handled carefully to avoid air-drying, which would otherwise lead to agglomeration of the Si lamellae. The drying was performed by transferring the samples from water to absolute ethanol, and subsequently to hexamethyldisilazane (HMDS, 100%) for 2 min. Finally, the samples were allowed to air-dry under the chemical hood for 10 min.

*Electrodeposition:* To realize the Au electrodeposition, a conductive seed layer was deposited by means of physical vapor deposition (Lesker PVD75 Sputter Coater system). First, a Cr adhesive layer was applied (DC 500 W, 5 mTorr) with deposition rate of 11 nm min$^{-1}$. Next, Au was deposited (DC 50 W, 3.75 mTorr) with the rate of 6.5 nm min$^{-1}$. As prepared samples were electroplated in NB Semiplate Au 100 (Arsenic-based, MicroChemicals) by using the current density of 0.112 mA cm$^{-2}$ at 30 °C. The current was applied using a PGSTAT204 (MetrOhm) supported by the software NOVA.

*Material characterization:* Cross-section and top view scanning electron microscopy (SEM) images were taken by Carl Zeiss XB1540 SEM and SmartSEM software (equipped with tilt correction) at 2–10 kV operating voltage. Prior to the imaging, samples were manually cut into smaller sections and some samples were sputter-coated with Au. Note, the manual cut can induce defects, whose origin, however, is apparent when analyzing the micrographs. ImageJ was used for statistical analysis of the nanostructure dimensions such as height and widths (with 50 quantities measured).

The pitch and duty cycle of the gratings were measured using top view images at SEM (e.g., Figure S1c) by our developed Python script. The script acquired and binarized the image finding the edges of the gratings using Canny edge detection. The average period and width of the grating was then measured from the middle points between the detected edges. The image magnification was chosen as high as possible to have a consistently low pixel/nm ratio to reduce the inaccuracy of the algorithm.

*Electrochemical deposition characterization:* The gold deposit was observed by the SEM and showed a compact structure growing from the sidewalls inward (Figure 4). After filling up the trenches, a gold overgrowth atop occurs. To control the latter, the mass *m* of gold to be deposited to completely fill the trenches was predicted using Faraday's law:

$$m = \frac{I\,t\,M_w}{z\,F}, \text{(Eq. 2)}$$

Where $I$ is the direct current applied, $t$ is the total plating time, $M_w$ is gold molecular weight, $z$ is valence number, and $F$ the Faraday's constant. For each sample, the surface area $A$ to immerse in solution was measured before the plating using ImageJ and its mass using a precision balance. Therefore:

$$\rho V = \frac{I\, t\, M_w}{z\, F}, \quad (Eq.\ 3)$$

$$\frac{\rho A h}{2} = \frac{I\, t\, M_w}{z\, F}, \quad (Eq.\ 4)$$

$$\frac{\rho h}{2} = \frac{I/_A\, t\, M_w}{z\, F} = \frac{J\, t\, M_w}{z\, F}, \quad (Eq.\ 5)$$

Where $\rho$ is the bulk gold density (19.3 $g\ cm^{-1}$), $V$ the volume to electroplate, $h$ is the height of the gratings, and the factor ½ accounts for the volume of the trenches to be plated (assuming a duty cycle of 50%). $J$ is the current density applied during the plating and it was kept constant to have the same deposition rate for every sample. Using this equation, plating time and deposited mass was estimated. The values of final mass, time and volume plated agreed with the predicted values.

The method is further validated by the fact that the trench over-plating consistently started after the time predicted for the trenches to be filled. The onset correlates with a negative shift of 0.01-0.02 V, clearly visible in the in-operando voltage profile of the software NOVA (see Figure S8g). The different environment where the gold grows, from a narrow trench to the top of the structure, changes the ions concentration at the metal-electrolyte interface, thus negatively changing the voltage needed to maintain the set current density. Consequently, the negative shift can act as a probe to calibrate the metallization to occur only in between high aspect ratio structures without overfilling.

*X-ray performance:* The diffractive performance of one of the gratings was tested at the I13-a beamline of the Diamond Light Source (Didcot, Oxfordshire, UK) using a hard X-ray monochromatic beam of photon energy 12.2 keV. This energy was selected as it is expected to result in a π phase modulation according to the design specifications of the grating (414 nm-period and a 3.2 ± 0.2 µm height, trenches filled with gold at 87% of its nominal density).

The collimated beam size was defined by using a set of X-ray slits upstream the grating to be 440 µm to prevent the diffracted orders of the grating from overlapping at the detector. The detector (a PCO edge scientific CMOS coupled to a scintillator via an objective, resulting in an effective pixel size of 0.8 µm) was placed at 3.3 m from the grating. This arrangement allowed resolution of the $0^{th}$ and the $\pm 1^{st}$ diffraction orders which were separated by 840 µm at the detector. To image the higher orders, the detector was moved laterally in both directions. To merge the images, an intensity profile of each image was taken, with the overlapping features identified and averaged prior to the merging. Images of the unperturbed beam i.e., without the grating ($I_0$) and with the beam off ($I_{dark}$) were collected to apply flat and dark-field corrections to the grating image $I$ as per equation (6):

$$\tilde{I} = \frac{(I - I_{dark})}{(I_0 - I_{dark})}, \text{ (Eq. 6)}$$

Where $\tilde{I}$ is the corrected image (a profile from which is plotted in Figure 5) used to calculate the diffraction order powers.

*Design of gratings and validation simulations:* The fabrication method allows for gratings to be designed to achieve a specified phase modulation between the Au filled trench, and the Si lamellae at a given X-ray energy. For example, if a phase modulation of $\phi$ is desired at an energy of $E$, then the necessary trench height, $L$, can be determined by:

$$L = \frac{\phi \lambda_E}{2\pi(\delta_{Au}(\rho_{Au},E) - \delta_{Si}(\rho_{Si},E))}, \text{ (Eq. 7)}$$

where $\lambda_E$ is the wavelength at energy $E$, and $\delta_{Au}$, $\delta_{Si}$, are the material deltas which themselves are a function of the material densities, $\rho_{Au}$, $\rho_{Si}$, and the energy $E$.

In house beam propagation software[37] was used to perform wave optical simulations of the proposed gratings, and of the experimental setup. The simulations explicitly include the grating structure, effective source to grating distances, as well as grating to detector distances. The grating used in the experiment at the I13 beamline of the Diamond Light Source (Didcot, Oxfordshire, UK), was designed to attain a $\phi = \pi$ phase modulation at an energy of 12.2keV. Assuming a 100% density of Au in the trenches, this requires a grating height of $L$ = 3.2 µm. After comparison of the experimental results to simulations, it was determined that the density of Au in the trenches is likely 87% of nominal (bulk) Au density, and thus the achieved phase modulation of the 3.2 µm grating is less than π. This revised estimate of the achieved Au density can be used in the equation above to improve the accuracy of the target phase modulation.

**Supporting Information**

Supporting Information includes 1) Methods section, 2) Supporting Figures S1-S10 detailing fabrication, wetting properties, and demonstrating collapse behavior, and 3) Supporting Text S1and S2 describing theory of the collapse behavior and electroplating details, respectively.

**Acknowledgments**

This work is funded by EPSRC (Grant EP/T005408/1). GK is supported by the Slovenian Research Agency (ARRS): research program P1-0192 and research project J1-3006. AO is supported by the Royal Academy of Engineering under the "Chairs in Emerging Technologies" scheme (Grant

CiET1819/2/78). PM is supported by a Royal Society University Research Fellowship (URF\R\191036). The authors gratefully acknowledge the provision of beam times MG28831-1 and MG28831-2 on I13-1 at the Diamond Light Source, and the assistance of the beamline scientists there. The authors acknowledge the assistance of the technical team in the London Centre for Nanotechnology (LCN): Steve Etienne, Vijayalakshmi Krishnan, Lorella Rossi, Rohit Khanna, and Suguo Huo.**Conflict of Interest**

The authors declare no competing financial interest.

**References**

[1]   M. S. Schmidt, J. Hübner, A. Boisen, *Advanced Materials* **2012**, *24*, OP11.

[2]   A. P. Yuda, P. Y. E. Koraag, F. Iskandar, H. S. Wasisto, A. Sumboja, *J Mater Chem A Mater* **2021**, *9*, 18906.

[3]   J. He, Z. Yang, P. Liu, S. Wu, P. Gao, M. Wang, S. Zhou, X. Li, H. Cao, J. Ye, *Adv Energy Mater* **2016**, *6*, 1501793.

[4]   P. Lecointre, S. Laney, M. Michalska, T. Li, A. Tanguy, I. Papakonstantinou, D. Quéré, *Nat Commun* **2021**, *12*, 3458.

[5]   M. Michalska, S. K. Laney, T. Li, M. K. Tiwari, I. P. Parkin, I. Papakonstantinou, *Nanoscale* **2022**, *14*, 1847.

[6]   M. Michalska, F. Gambacorta, R. Divan, I. S. Aranson, A. Sokolov, P. Noirot, P. D. Laible, *Nanoscale* **2018**, *10*, 6639.

[7]   L. Romano, M. Kagias, J. Vila-Comamala, K. Jefimovs, L.-T. Tseng, V. A. Guzenko, M. Stampanoni, *Nanoscale Horiz* **2020**, *5*, 869.

[8]   C. Chang, A. Sakdinawat, *Nat Commun* **2014**, *5*, 4243.

[9]   S. K. Lynch, C. Liu, N. Y. Morgan, X. Xiao, A. A. Gomella, D. Mazilu, E. E. Bennett, L. Assoufid, F. de Carlo, H. Wen, *Journal of Micromechanics and Microengineering* **2012**, *22*, 105007.

[10]   H. Miao, A. Panna, A. A. Gomella, E. E. Bennett, S. Znati, L. Chen, H. Wen, *Nat Phys* **2016**, *12*, 830.25

[11]     L. Romano, M. Stampanoni, *Micromachines (Basel)* **2020**, *11*, 589.

[12]     X. Li, P. W. Bohn, *Appl Phys Lett* **2000**, *77*, 2572.

[13]     C. Chiappini, In *Handbook of Porous Silicon* (Ed.: Canham, L.), Springer International Publishing, Cham, **2014**, pp. 171–186.

[14]     A. Mallavarapu, P. Ajay, S. v. Sreenivasan, *Nano Lett* **2020**, *20*, 7896.

[15]     A. S. Togonal, L. He, P. Roca I Cabarrocas, Rusli, *Langmuir* **2014**, *30*, 10290.

[16]     A. E. Hollowell, C. L. Arrington, P. Finnegan, K. Musick, P. Resnick, S. Volk, A. L. Dagel, *Mater Sci Semicond Process* **2019**, *92*, 86.

[17]     F. Koch, F. Marschall, J. Meiser, O. Márkus, A. Faisal, T. Schröter, P. Meyer, D. Kunka, A. Last, J. Mohr, *Journal of Micromechanics and Microengineering* **2015**, *25*, 075015.

[18]     L. Romano, J. Vila-Comamala, K. Jefimovs, M. Stampanoni, *Adv Eng Mater* **2020**, *22*, 2000258.

[19]     C. N. Nanev, E. Saridakis, N. E. Chayen, *Sci Rep* **2017**, *7*, 35821.

[20]     H. Zhou, F. G. Shi, B. Zhao, *Microelectronics J* **2003**, *34*, 259.

[21]     D. Wheeler, D. Josell, T. P. Moffat, *J Electrochem Soc* **2003**, *150*, C302.

[22]     D. Josell, T. P. Moffat, *J Electrochem Soc* **2017**, *164*, D327.

[23]     H. Miao, A. A. Gomella, N. Chedid, L. Chen, H. Wen, *Nano Lett* **2014**, *14*, 3453.

[24]     J. Vila-Comamala, S. Gorelick, E. Färm, C. M. Kewish, A. Diaz, R. Barrett, V. A. Guzenko, M. Ritala, C. David, *Opt Express* **2011**, *19*, 175.

[25]     S. K. Laney, T. Li, M. Michalska, F. Ramirez, M. Portnoi, J. Oh, M. K. Tiwari, I. G. Thayne, I. P. Parkin, I. Papakonstantinou, *ACS Nano* **2020**, *14*, 12091.

[26]     F. Braet, R. De Zanger, E. Wisse, *J Microsc* **1997**, *186*, 84.

[27]     M. Michalska, S. K. Laney, T. Li, M. Portnoi, N. Mordan, E. Allan, M. K. Tiwari, I. P. Parkin, I. Papakonstantinou, *Advanced Materials* **2021**, 2102175.

[28]     S. Si, M. Hoffmann, *Microelectron Eng* **2018**, *197*, 39.

[29]     O. J. Hildreth, W. Lin, C. P. Wong, *ACS Nano* **2009**, *3*, 4033.

[30]     D. Gräf, M. Grundner, R. Schulz, L. Mühlhoff, *J Appl Phys* **1990**, *68*, 5155.

[31]     Stephen Timoshenko, James M. Gere, *Theory of elastic stability*, 2nd ed., Dover ed., Dover Publications, Mineola, N.Y., 2009.

[32]     C. Y. Hui, A. Jagota, Y. Y. Lin, E. J. Kramer, *Langmuir* **2002**, *18*, 1394.

[33]   S. Znati, N. Chedid, H. Miao, L. Chen, E. E. Bennett, H. Wen, *J Surf Eng Mater Adv Technol* **2015**, *05*, 207.

[34]   A. Pei, G. Zheng, F. Shi, Y. Li, Y. Cui, *Nano Lett* **2017**, *17*, 1132.

[35]   Y. Zheng, C. Li, H. Hu, S. Huang, Z. Liu, H. Wang, *Jpn J Appl Phys* **2021**, *60*, 035003.

[36]   P. Lianto, S. Yu, J. Wu, C. V. Thompson, W. K. Choi, *Nanoscale* **2012**, *4*, 7532.

[37]   P. R. T. Munro, *Journal of the Optical Society of America A* **2019**, *36*, 1197.

Phase-based X-ray imaging techniques achieve unmatched resolution for low-Z objects but often rely on diffracting elements requiring challenging fabrication processes. Here, a key facile innovation to post-MACE drying made submicron-period gratings with high aspect ratio (>40) feasible, being compatible with inexpensive lithography and dense electroplating out-of-cleanroom processes. The gratings' performance was tested in a synchrotron facility showing excellent diffracting abilities.

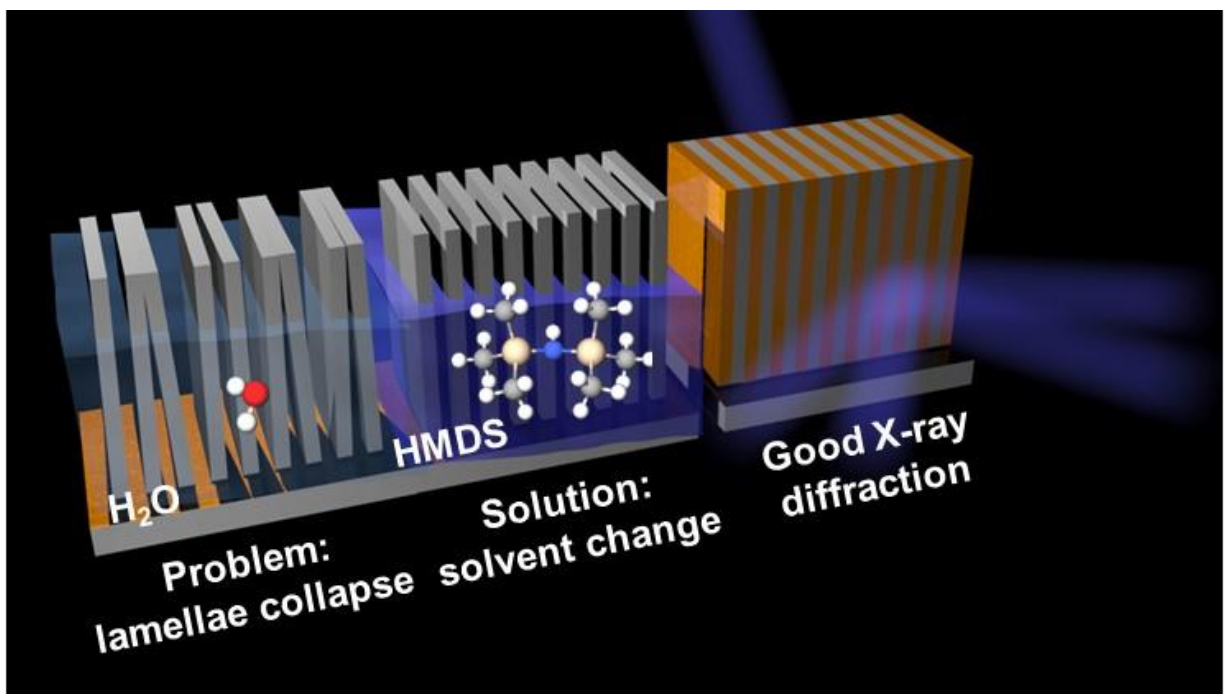

**Supporting Figures**

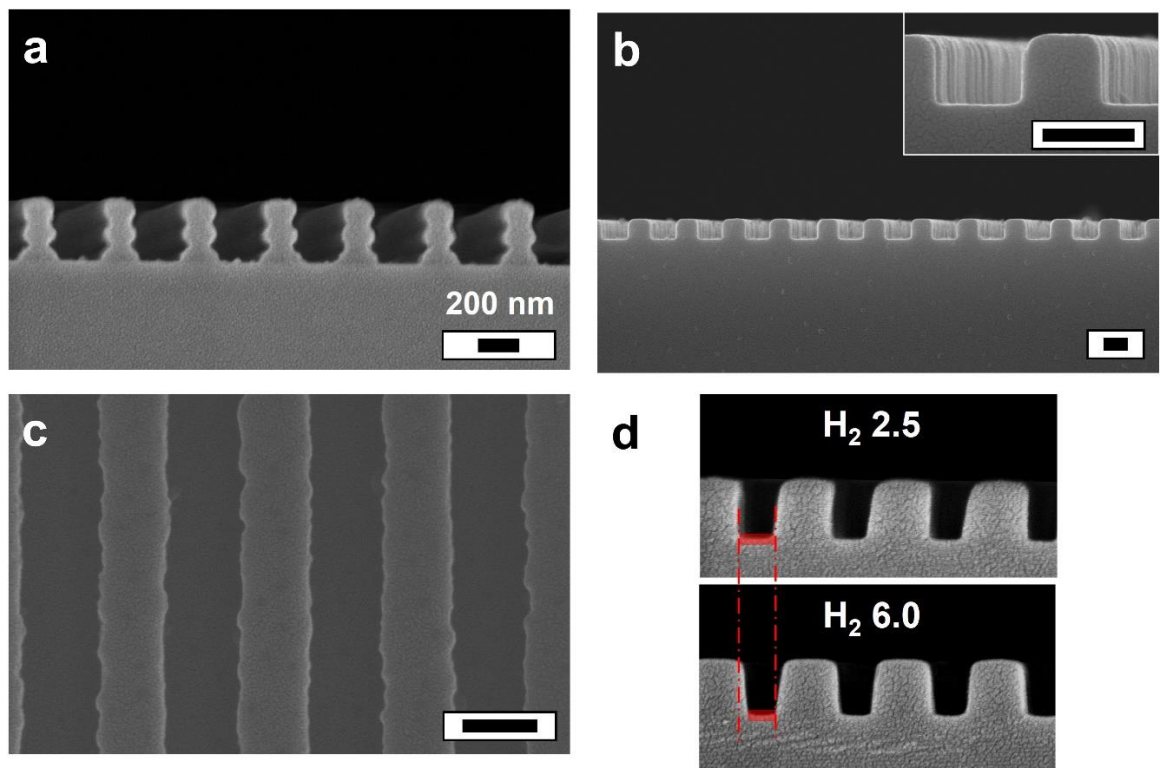

**Figure S1.** Master preparation. The nanograting lines were generated by means of LIL in a photoresist layer deposited on Si/SiO$_2$ wafer. Subsequently, the pattern was transferred into the glass layer by RIE. (a) Cross-sectional scanning electron microscopy (SEM) image of the photoresist pattern after development. (b) Cross-sectional and (c) top view SEM images of the master after the RIE, where the grating is glass, and the substrate is silicon. Glass etching was performed in CHF$_3$/Ar plasma with or without addition of H$_2$. (d) SEM images of two masters with different duty cycles (=grating width/pitch). The higher H$_2$ flux, the greater the width and thus, the duty cycle. Scale bars are 200 nm.

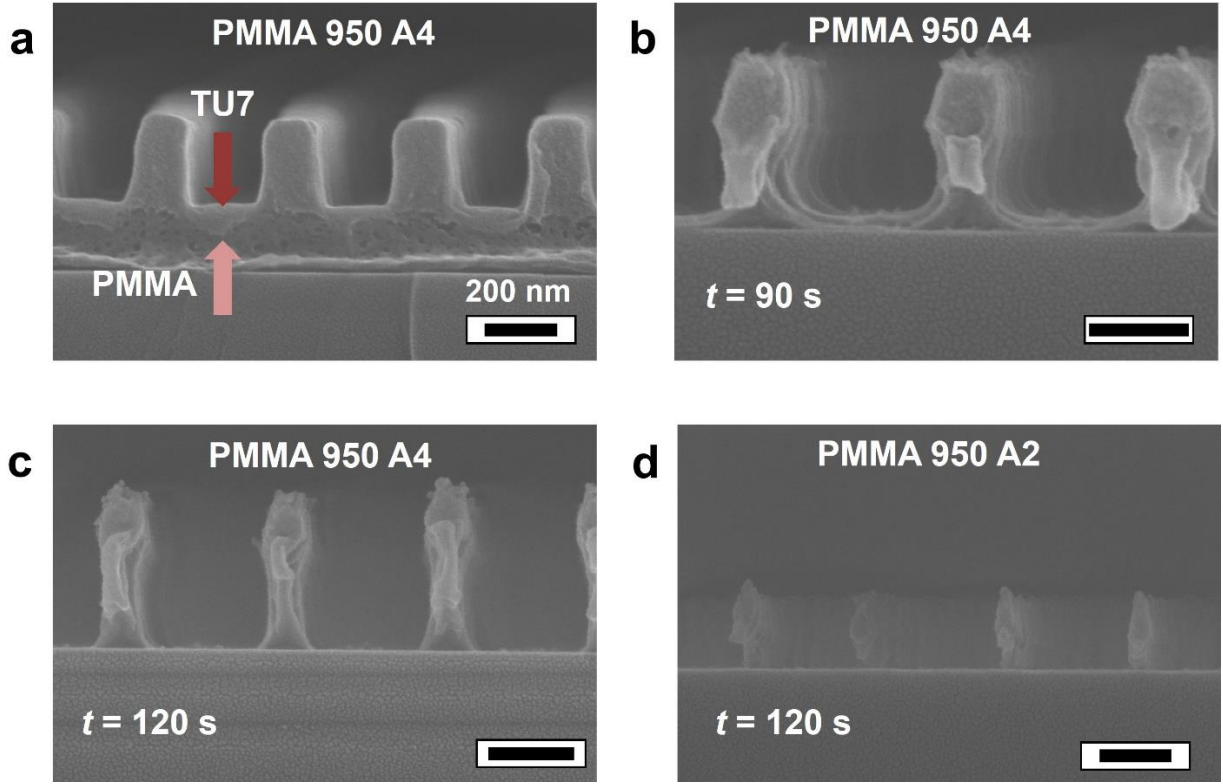

**Figure S2.** Optimization of the bi-layer lift-off process via nanoimprint lithography. SEM cross-section images of (a) PMMA (950 A4) below STU-7 stamped with master with linear grating pattern, and (b-c) various times *t* = 90 and 120 s of oxygen breakthrough . (d) SEM cross-section image of the sample with a thinner PMMA layer (905 A2) subjected to 120 s of the oxygen breakthough. Both images (c) and (d) show undercut but alos a drastic change in duty cycle.

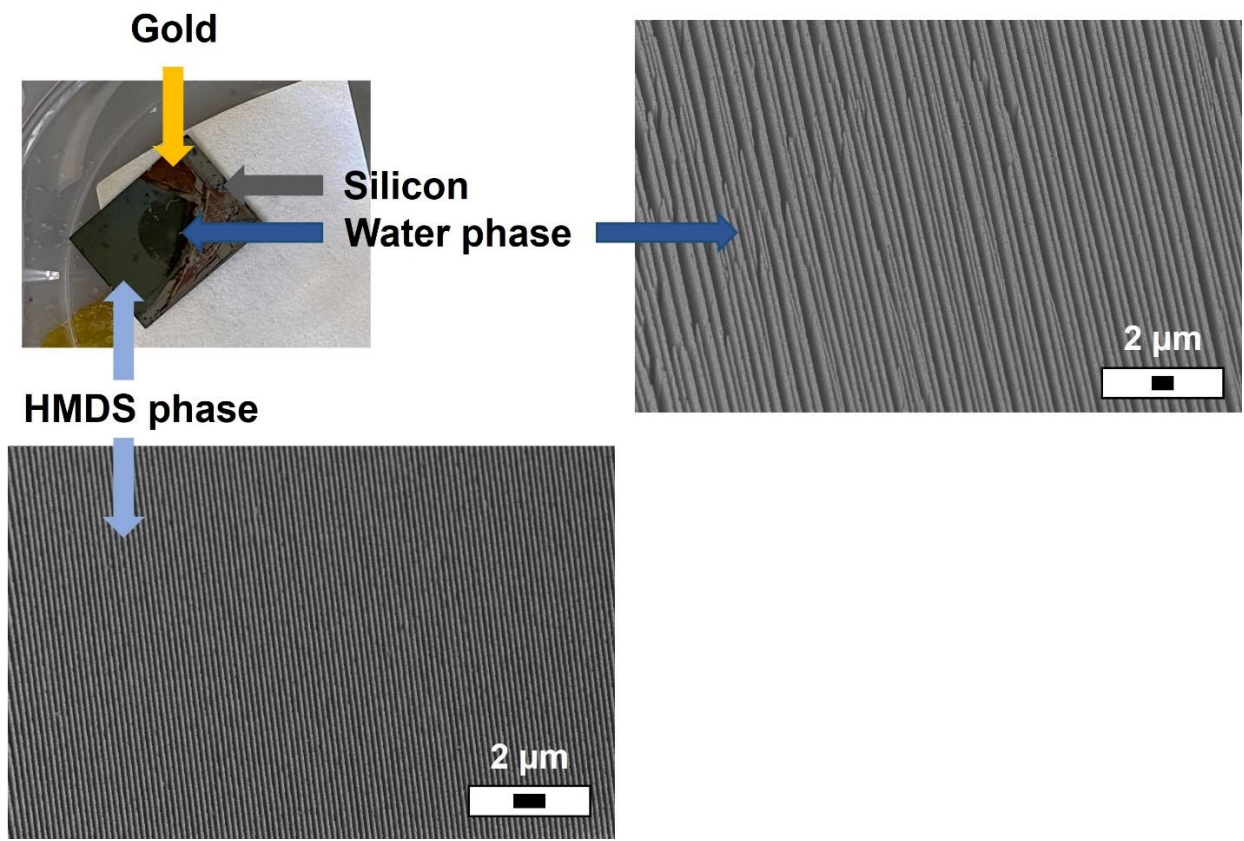

**Figure S3.** HMDS and water phase separation on the nanograting. A digital photograph of the silicon grating fabricated via MACE, rinsed in water, and transferred directly to 100% HMDS solution for 2 min. Subsequently, the sample was removed and left to air-dry. During the drying process, a clear phase separation appeared with the water phase being pushed into a corner (darker area, dark blue arrows) and the rest of the grating drying within seconds (lighter area, light blue arrows). The SEM top images reveal undistorted and distorted patterns. Scale bars 2 µm.

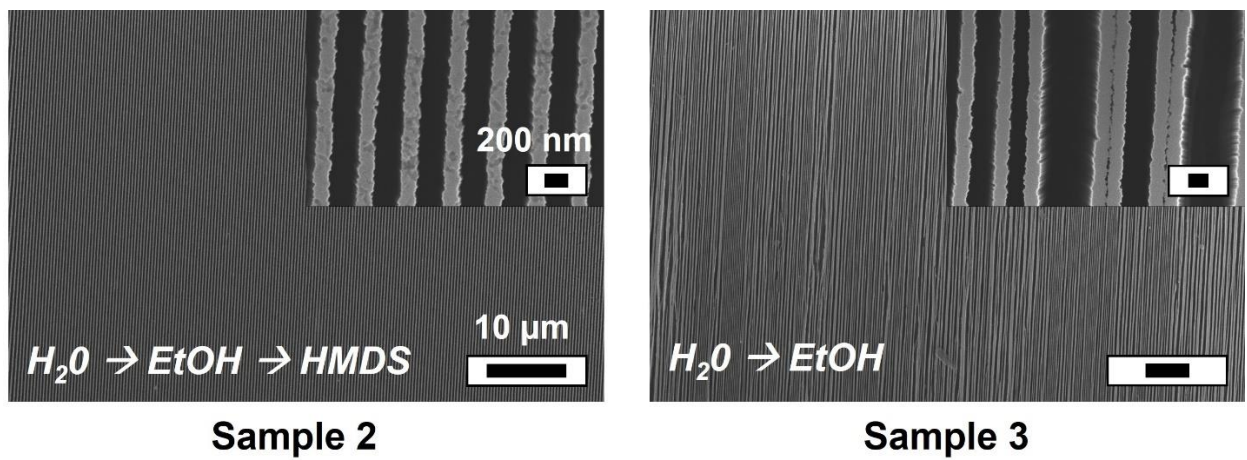

**Figure S4.** A comparison of two drying scenarios from low surface tension liquids. SEM top view images of nanogratings dried from water through ethanol and HMDS (Sample 2) and only ethanol (Sample 3). Scale bars are 10 µm. The inset show magnified views with the defect-free lamellae and the collapsed ones. Scale bars of the insets are 200 nm.

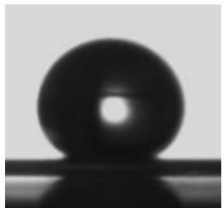 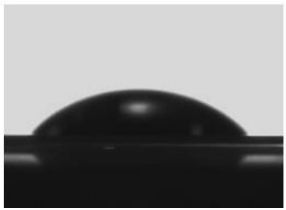 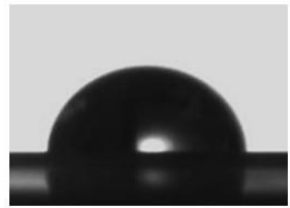

**Figure S5.** A comparison of the water contact angles (WCA) measured after the MACE process for the flat and patterned silicon (grating). Left: Water is in a nearly superhydrophobic state (WCA = 147°) on the grating after drying in HMDS. Middle: Flat silicon (no pattern) after the etching and drying directly from water yields a less hydrophilic surface (WCA = 48°) in comparison to the starting superhydrophilic material, where water spreads instantly (WCA < 10°). Right: Flat silicon after the etching and drying in HMDS yields even less hydrophilic surface (WCA = 87°), indicating the HMDS bonding.

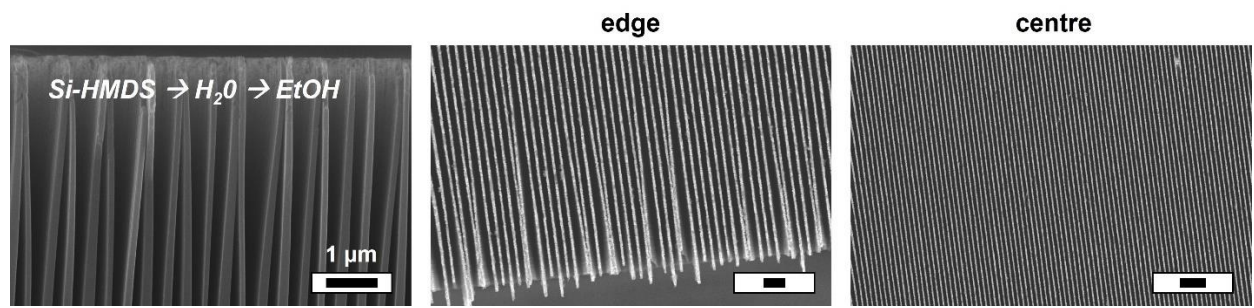

**Figure S6.** Inducing collapse behavior on HMDS-coated silicon gratings. SEM cross section (left) and top views (middle, right) showing collapsed lamellae at the edge and free standing in the middle of the ca. 1 cm$^2$ sample. The sample was previously dried via HMDS step and subsequently, it was sonicated in water, followed by ethanol. Scale bars are 1 µm.

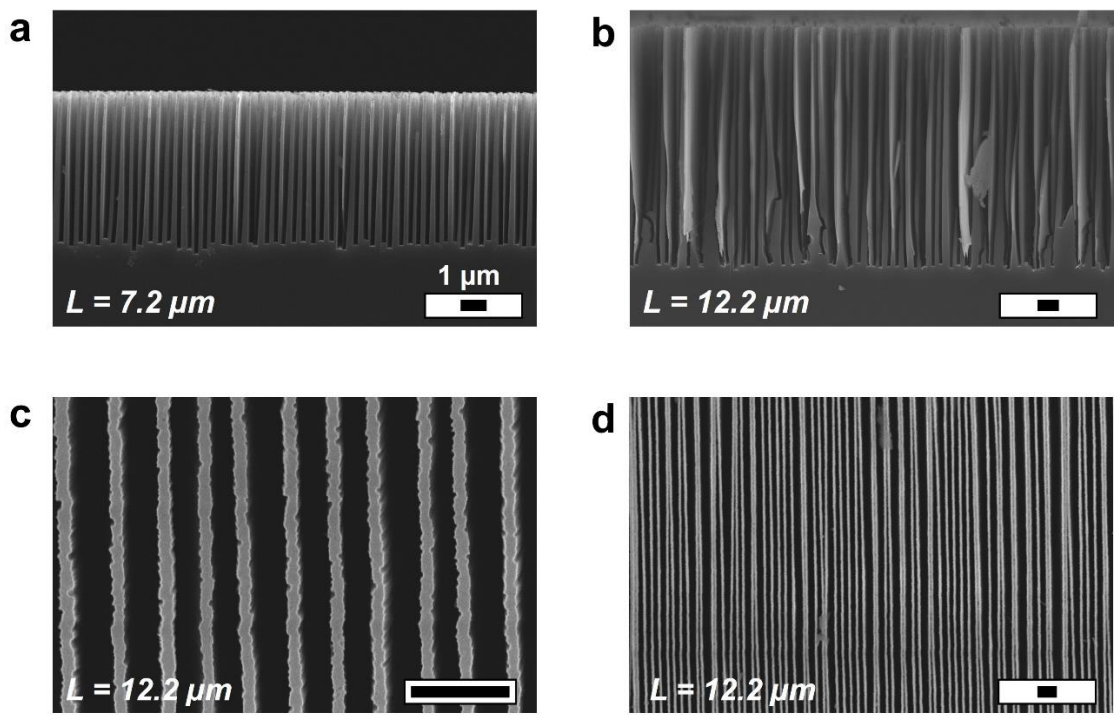

**Figure S7.** Experimental limits of the aspect ratio of nanogratings after HMDS-mediated drying. SEM cross section images of the nanograting etched for (a) 2h 40 min and (b) 4h 30 min yielding 7.2 and 12.2 μm tall structures, respectively. The corresponding aspect ratios are 43 and 117. (c-d) Top view SEM images of the 12.2 μm tall grating showing defect-free regions in (c) and some lamellae collapsed in (d). The lack of verticality as seen in (b) is likely a cause of the collapse, which implies the attainable aspect ratio are >100.

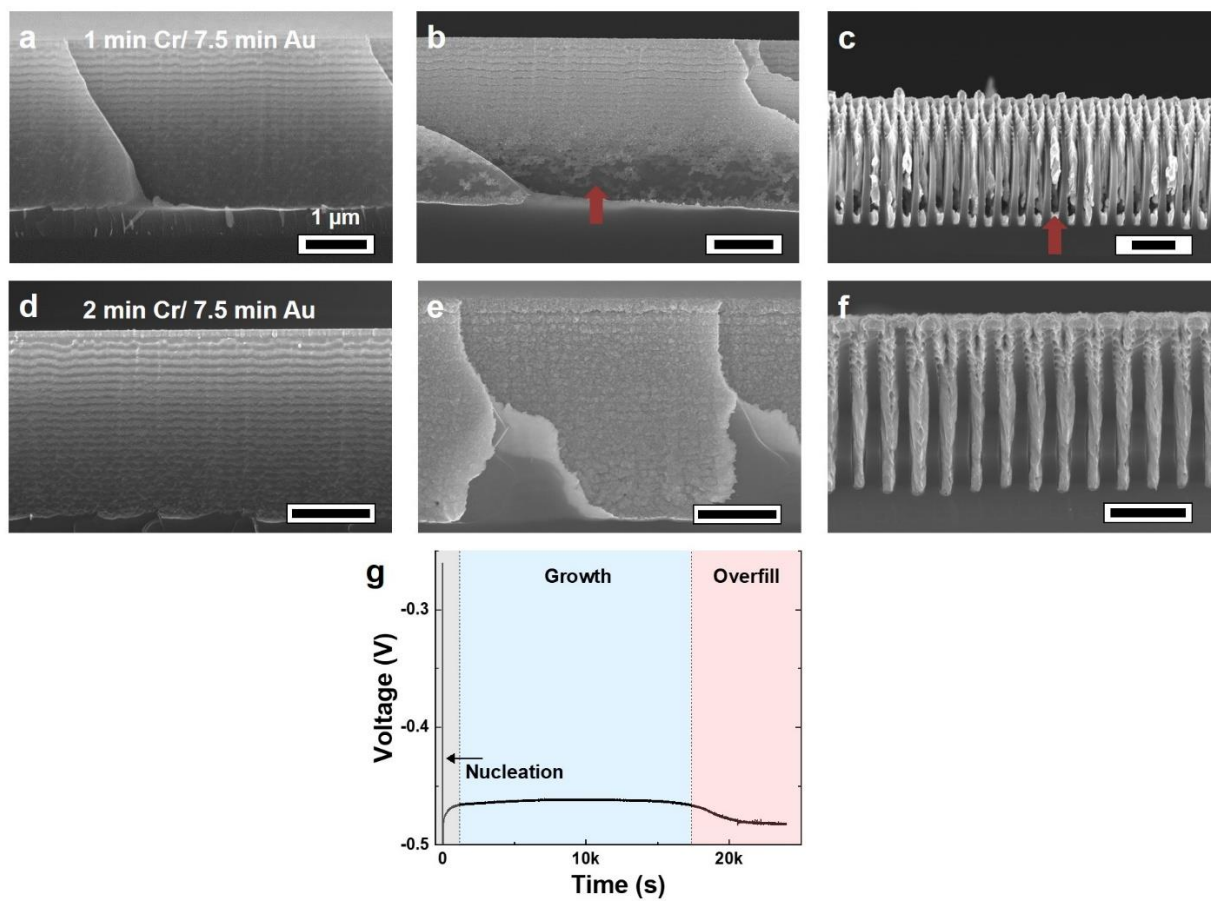

**Figure S8.** Effect of thickness of an adhesive layer during the subsequent gold sputtering on high-aspect ratio nanogratings. SEM cross section images showing a Cr/Au seed layer (a-b, d-e) sputtered onto silicon gratings and the results of gold electroplating (c, f). Chromium serves as the adhesive layer and was sputtered for 1 min (a-c) and 2 min (d-f), yielding ca. 10 and 20 nm thickness, respectively. The gratings were fabricated by deep reactive ion etching (DRIE) process in this instance. Scale bars are 1 µm. (g) A voltage profile of an electroplating process on 5.3 µm-tall grating showing an initial nucleation step (grey), a growth phase inside the trenches (blue), and the top layer formation at the end (overfill, pink).

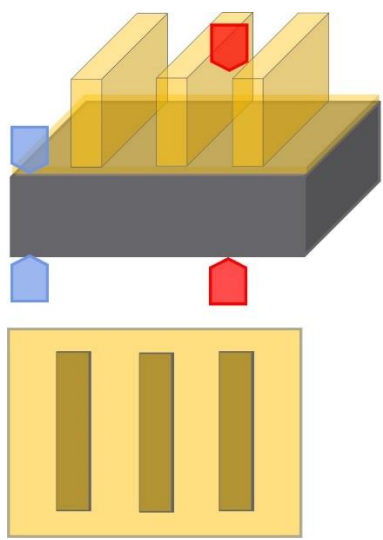

**Figure S9.** Grating design. Schematic tilted side and top views of the gratings surrounded by an Au-coated frame. The frame is etched down with the gratings and coated so that the electrode during electrochemical deposition has a contact at the bottom (blue arrows). This is to ensure conductivity across the sample and at the bottom of the trenches that an electrical contact at the top of the gratings (red arrows) would not guarantee for poorly conductive seed layers. It also prevents a mechanical damage to the grating during its handling.

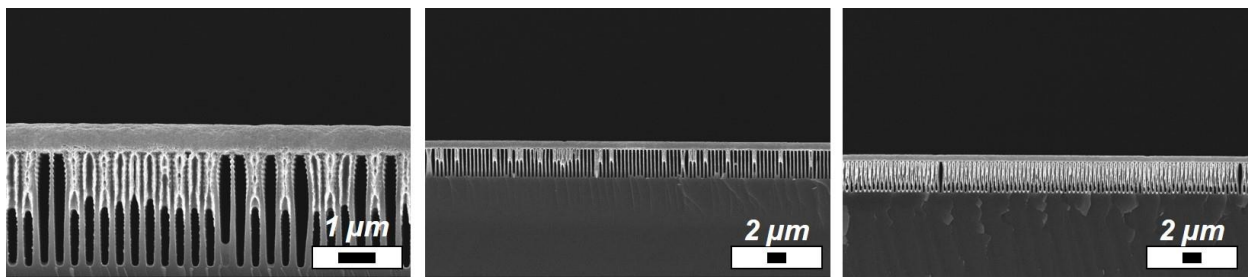

**Figure S10.** The effect of grating design on electroplating. The SEM images show unframed samples prepared by deep reactive ion etching process after Au electroplating process. We observe a limited/none growth at the bottom part of the trenches.

*Supporting Text*

**Text S1 – Collapse behavior theory**

This section provides the equations we used for different estimations of the maximal lamella height, $L_{max}$, before a collapse occurs. Historically, the first consideration was given to self-buckling, a collapse initiated solely by gravity acting on the lamella. Considering a cantilever prismatic Euler-Bernoulli beam standing upright, we can calculate $L_{max}$ for self-buckling, $L_{maxSB}$, using the well-known Greenhill result:

$$L_{maxSB} = \sqrt[3]{7.8373 \frac{E\,I}{\rho g A}}, \text{ (Eq. S1)}$$

where $E$ is the Young's modulus for lamella, $I$ is the second moment of area (area moment of inertia), $\rho$ is the lamella material density, $g$ is the gravitational acceleration, and $A$ is the beam cross-sectional area. We obtain the lowest $L_{maxSB}$ estimate, when $I$ is minimal, i.e. the buckling occurs around the axis running along $d$ (with deflection in the $b$ direction). This is assumed for all the calculations that follow. Eq. (S1) specific for our case is expressed as:

$$L_{maxSB} = \sqrt[3]{7.8373 \frac{E\,b^2}{12\,\rho\,g}}, \text{ (Eq. S2)}$$

and for our experimental parameters yields 5.3 mm.

With self-buckling ruled out, we now consider scenarios, where a force bends two lamellae until they come into contact. Once the sides of the lamellae touch, a force arising from the surface interaction appears. This force drives the lamella to increase contact area until the increasingly deformed bottom part of the lamella balances it out. Following the result from Hui *at al.*[1] and using our notation for geometrical parameters we obtain an $L_{max}$ for surface energy interaction, $L_{maxSE}$:

$$L_{maxSE} = \sqrt[4]{\frac{3\,E\,b^3\,(p-b)^2}{4\,(1-\nu^2)\,\gamma_S}}, \text{ (Eq. S3)}$$

where $\nu$ is the Poisson's ratio and $\gamma_S$ is the surface energy of the lamella material. For our parameters $L_{maxSE}$ = 2 μm. The computed $L_{maxSE}$ according to Eq. (8) can be larger, if taken into account, that the surfaces are also equally charged.

First, we acknowledge that any force involved in bending will need to counteract the elastic restoring force for a cantilever beam deflection by half the pitch (*p/2*). Experimentally, we know the critical step that initiates the collapse is the drying process. Therefore, it is reasonable to assume that capillary forces are responsible for bending the lamellae together. When considering the capillary force for our geometry, we make the simplification of two infinite parallel lamellae and solve the Laplace equation to obtain the force per unit length.[2] In order to get an analytical expression, there are further assumptions necessary. Namely, we assume that the meniscus height between the lamellae is constant and pinned. Because this capillary force increases with pinning height, we set it to be equal to lamella height to obtain $L_{max}$ for capillary force, $L_{maxCF}$:

$$L_{maxCF} = \sqrt[5]{\frac{E\, b^3\, p\, (\sinh(q\, p))^2}{8\, \gamma\, q^2\, (\cosh(q\, p) - 1)}}, \text{(Eq. S4)}$$

$$q = \sqrt{\frac{\gamma}{\Delta \rho\, g}}, \text{(Eq. S5)}$$

where $\gamma$ is surface tension, $q$ is capillary length and $\Delta\rho$ is the density difference between the two fluids on either side of the meniscus. In our system, the final solvent in the drying process is HMDS exposed to air, and we can use Eq. (S4) and (S5) to calculate $L_{maxCF}$ = 107 µm.

Here we present the elastic restoring force *F* comparison for the deflection of the entire grate to the deflection of corner only. The basic equation comes from the theory of Euler-Bernoulli cantilever beams for small deflections $\delta$ and reads:

$$\frac{d^2\delta}{dx^2} = -\frac{12\, F\, x}{E\, b^3 d}, \text{(Eq. S6) and}$$

$$\frac{d^2\delta}{dx^2} = -\frac{6\, F\, \sin(2\alpha)}{E\, b^3}, \text{(Eq. S7)}$$

where we choose *x* to be the deflection direction for a complete rectangular grate (Eq. S6) and its corner only (Eq. S7). See main text Figure 3e for definition of α. We obtain the force *F* for each of the two cases and establish the condition from the main text by asking when *F* from (S6) is larger than F from (S7).

**Text S2 – Electroplating: electrical contact**

A void-free, homogenous filling of the high-aspect ratio nanotrenches with gold is a non-trivial task. Here, we address this challenge with two combined efforts: 1) a continuous seed layer of Cr/Au sputtered along the sidewalls to promote a homogeneous nucleation, and 2) a particular design of protruding gratings to increase the reactants' accessibility inside the trenches (Figure S9).

A ~5 mm margin (frame) was left unexposed during LIL process, resulting in protruding grating from the substrate after the pattern transfer via etching. Here, the rationale is that such a protruding design not only adds a lateral diffusion pathway inside the trenches, increasing plating homogeneity across the sample, but also ensures an electrical contact with the bottom of the trenches, which drives the growth towards the bottom, otherwise hindered due to low diffusion. In addition, such design allows for a safer handling of the samples.

We performed electroplating on framed/unframed design. Whilst nanogratings fabricated via MACE were plated equally well regardless the design, the choice of the framed one proved to be essential for the successful gold deposition on the samples prepared via deep reactive ion etching (DRIE). Although this latter is a part of an ongoing project, we showcase it here to flag an importance for this design consideration. The unframed DRIE samples showed a limited/none growth at the bottom part of the trenches, as presented in Figure S10, as well as marked heterogeneity across the sample surface area.

These results suggest that a better electrical contact along the sidewalls is established in the MACE-samples in comparison to the DRIE-samples. The poorer conductivity along the trenches in the gratings etched by DRIE may be attributed to the scalloping introduced by the etching method.

**References**

[1]    C. Y. Hui, A. Jagota, Y. Y. Lin, E. J. Kramer, *Langmuir* **2002**, *18*, 1394.

[2]    V. N. Paunov, *Langmuir* **1998**, *14*, 5088.